\documentstyle[12pt]{article}

\def\stackunder#1#2{\mathrel{\mathop{#2}\limits_{#1}}}%

\begin{document}

\input{epsf}

\def\beq{\begin{equation}}
\def\eq{\end{equation}}
\def\bar{\begin{eqnarray}}
\def\ear{\end{eqnarray}}
\def\bars{\begin{eqnarray*}}
\def\ears{\end{eqnarray*}}
\def\ov{\overline}
\def\ot{\otimes}

\def\deb{\frac{1}{2}}
\newcommand{\dd}{\mbox{$\Delta$}}
\newcommand{\af}{\mbox{$\alpha$}}
\newcommand{\be}{\mbox{$\beta$}}
\newcommand{\la}{\mbox{$\lambda$}}
\newcommand{\gh}{\mbox{$\gamma$}}
\newcommand{\ep}{\mbox{$\epsilon$}}
\newcommand{\vep}{\mbox{$\varepsilon$}}
\newcommand{\de}{\mbox{$\frac{1}{2}$}}
\newcommand{\th}{\mbox{$\frac{1}{3}$}}
\newcommand{\qa}{\mbox{$\frac{1}{4}$}}
\newcommand{\sx}{\mbox{$\frac{1}{6}$}}
\newcommand{\vg}{\mbox{$\frac{1}{24}$}}
\newcommand{\tde}{\mbox{$\frac{3}{2}$}}

\newcommand{\np}{{\it Nucl. Phys.}}
\newcommand{\pl}{{\it Phys. Lett.}}
\newcommand{\prl}{{\it Phys. Rev. Lett.}}
\newcommand{\cmp}{{\it Commun. Math. Phys.}}
\newcommand{\jmp}{{\it J. Math. Phys.}}
\newcommand{\jpamg}{{\it J. Phys. {\bf A}: Math. Gen.}}
\newcommand{\lmp}{{\it Lett. Math. Phys.}}
\newcommand{\ptp}{{\it Prog. Theor. Phys.}}

\title{Baxter Equation for the QCD Odderon}
\author{{\bf Ziad Maassarani}\\
 \\
{\bf Samuel Wallon}\\
 \\ 
{\small CEA, Service de Physique Th\'eorique, CE-Saclay}\\
{\small F-91191 Gif-sur-Yvette  Cedex, FRANCE}}
\date{}
\maketitle

\begin{abstract}
The Hamiltonian derived by Bartels, Kwiecinski and
Praszalowicz for the study of high-energy QCD in the
generalized logarithmic approximation was found to correspond to the
Hamiltonian of an integrable $XXX$ spin chain. We study the odderon
Hamiltonian corresponding to three sites by means of the Bethe Ansatz
approach. 
We rewrite the Baxter equation, and consequently the Bethe Ansatz equations,
as a linear triangular system.
We derive a new expression for the eigenvectors and the
eigenvalues, and discuss the quantization of the conserved quantities.
\end{abstract}
\vspace{7cm}
\noindent
\hspace{1cm}June 1995\hfill\\
\hspace*{1cm}T95/081\hfill\\
\hspace*{1cm}hep-th/9507056
\thispagestyle{empty}
\newpage
\setcounter{page}{1}

\section{Introduction}\label{intro}

The behaviour of hadronic scattering amplitudes at high-energies
for fixed transferred momentum $t$ is, with confinement, one of the most
interesting problems to be solved in the strong interaction field.

The discovery of QCD gave us the theory for studying this problem. In the
framework of perturbative QCD, the resummation of leading logarithmic
amplitudes was performed \cite{bfkl} and gave the so-called ``perturbative
Pomeron'', which violates the Froissart bound \cite{froissart} derived from the
analyticity and unitarity of the $S$-matrix. This is why one needs to take
into account subleading terms, so as to restore unitarity.

It is possible to write down an action which takes into account all graphs
in the large $s$ limit \cite{lipeff}. But it is not easy to perform 
calculations in this framework. 
It turns out that one can consider a subclass of such
diagrams where one takes into account only the exchange of a fixed
number of reggeons in the $t$-channel, keeping only the dominant
contribution at each level \cite{bkp}. This model can be formulated as a
Schr\"{o}dinger equation with a two-body interaction Hamiltonian. This
Hamiltonian was later shown \cite{lipxxx,fk} 
to belong to the integrable hierarchy
of Hamiltonians of the Heisenberg $XXX$ spin-chain with 
$SL(2,${\bf C})-spin
zero. The Bethe Ansatz method was then applied in this framework to
diagonalize the Hamiltonian. In this paper we analyze the Baxter equation
and eigenvectors which arise in this context. In section (\ref{diff}) we derive
general results for the solution $Q_n(\lambda)$ of the Baxter equation for $n$
sites, and more particularly the $n=2$ and $n=3$ chains. In section
(\ref{poly}) we rewrite the Baxter equation in a new form, more appropriate to
the study of the polynomial solutions. And in section (\ref{vectors}) 
we give new expressions for the eigenvectors.

\section{The Reggeon Hamiltonian}\label{def}

After Fourier transform, in the two-dimensional impact parameter space, the
interaction Hamiltonian ${\cal H}_{jk}$ between two reggeons becomes the 
sum of a holomorphic piece and its anti-holomorphic counterpart.
The holomorphic
piece has the two equivalent representations \cite{lipham}:
\begin{eqnarray}
\label{hjk1}
H_{jk} &=&-P_j^{-1}\log \left( z_{jk}\right) P_j-P_k^{-1}\log \left(
z_{jk}\right) P_k-\log \left( P_jP_k\right) -2\gamma  \\
&=&-2\log (z_{jk})-z_{jk}\log \left( P_jP_k\right) z_{jk}^{-1}-2\gamma
\label{hjk2}
\end{eqnarray}
where
\[
z_{jk}=z_j-z_k \;,\;\; P_j=i\frac{\partial}{\partial z_j}\;,
\]
and $\gamma$ is the Euler constant.
There is also another equivalent form
\begin{equation}
H_{jk}=\stackunder{\ell =0}{\stackrel{\infty }{\sum }}\left( \frac{2\ell +1}{
\ell \left( \ell +1\right) -L_{jk}^2}-\frac 2{\ell +1}\right) \,,\;\;
L_{jk}^2=-z_{jk}^2\frac{\partial^2}{\partial z_j \partial z_k} \,.
\label{hjk3}
\end{equation}

The Hamiltonian (\ref{hjk3}) and its antiholomorphic counterpart
are clearly invariant under the conformal transformations \cite{liprev}
\begin{equation}
z_j\longrightarrow z_j^{^{\prime }}=\frac{az_j+b}{cz_j+d}\;\;,\;\; 
\overline{z}_j\longrightarrow \overline{z^{^{\prime }}}_j=\frac{\overline{a}
\overline{z}_j+\overline{b}}{\overline{c} \overline{z}_j+\overline{d}} 
\label{conft}
\end{equation}
with $ad-bc=\overline{a}\overline{d}-\overline{b}\overline{c}=1$

The complete reggeon Hamiltonian is given by
\[
{\cal H}_n=-\frac{\alpha_s}{2\pi }\sum_{n\geq j>k\geq 1}
\left( H_{jk}+\overline{H}_{jk}\right)t_j^a t_k^a 
\label{nc}
\]
where the $t_j^a$ are the color matrices for the $j^{\rm th}$ reggeon
and $\alpha_s$ is the strong coupling constant.

The eigenstates $\chi _{n,\left\{ q\right\} }$ are parametrized by a set of
quantum numbers $\left\{ q\right\} $ and the additional coordinates $\left(
z_0,\overline{z}_0\right) $ which correspond to the center of mass of the
compound Reggeon state. In the large-$N_c$ limit, one has
\beq
t_1^a t_2^a\longrightarrow -N_c\quad \hbox{for}\quad n=2\quad 
\quad t_j^at_k^a\longrightarrow -\frac{N_c}2\delta_{k,j+1}\quad 
\hbox{for}\quad n\geq 3\,,
\eq
and ${\cal H}_n$ becomes holomorphically separable:\footnote{
For $n=2$ and $n=3$, as one looks for color neutral reggeons, the
expression (\ref{nc}) is exact for any finite $N_c$. This is easily derived 
from the zero charge color condition $\stackunder{j=1}{\stackrel{n}{\sum }}
t_j^a=0 $ \cite{k}.} 
\beq
{\cal H}_n=\frac{\alpha _sN_c}{4\pi }\left( H_n+\overline{H}_n\right) =
\frac{\alpha_s N_c}{4\pi }\left( \sum_{j=1}^n H_{j,j+1}+
\sum_{j=1}^n\overline{H}_{j,j+1}\right) 
\eq
One can then look for eigenvectors in the form
\beq
\chi_{n,\left\{ q\right\} }\left( \{z_j,\overline{z}_j\};z_0,
\overline{z}_0\right) =\varphi_{n,\{ q\} }\left( \left\{ z_j\right\} ;
\overline{z}_0\right)\, \overline{\varphi }_{n,\{\overline{q}\} }\left(
\left\{ \overline{z}_j\right\} ;\overline{z}_0\right) , 
\eq
with
\beq
H_n\varphi_{n,\{ q\} } =\varepsilon _{n,\{ q\} }\varphi_{n,\{ q\} }\; ,\;\; 
\ov{H}_n\ov{\varphi}_{n,\{\ov{q}\} }=
\ov{\varepsilon}_{n,\{\ov{q}\} }
\ov{\varphi}_{n,\{\ov{q}\} }\,.
\eq
The eigenvectors also satisfy the conformal invariance property \cite{liprev}
\bar
\chi_{n,\{q\}}\left( \left\{ z_j,\overline{z}_j\right\} ;z_0,\overline{z}
_0\right) \rightarrow \chi _{n,\{q\}}\left( \left\{ z_j^{\prime },\overline{z
}_j^{\prime }\right\} ;z_0^{\prime },\overline{z}_0^{\prime }\right)
\nonumber\\ 
=\left( c z_0+d\right) ^{2h}\left( \overline{c}\overline{z}_0+
\overline{d}\right) ^{2h}\chi _{n,\{q\}}\left( \left\{ z_j,\overline{z}
_j\right\} ;z_0,\overline{z}_0\right)  \label{confo}
\ear
under the transformations (\ref{conft}). The conformal weights $h$ and $\ov{h}$
correspond to the principal series of $SL(2,${\bf C}) \cite{liprev}:
\[
h=\frac{1+m}2-i\nu \quad ,\quad \overline{h}=1-h^{*}=\frac{1-m}2-i\nu \,,
\]
with $m$ integer and $\nu$ real.

It was shown in \cite{lipxxx,fk} that $H_n$ is the nearest-neighbour
Hamiltonian of the spin zero XXX Heisenberg spin-chain with periodic
boundary conditions. The spin $s$ generators of $SL(2,${\bf C)} at
site $j$ are:

\[
S_j^{+}=z_j^2\partial _j-2 s z_j\;,\quad S^{-}=-\partial _j\quad ,\quad
S_j^3=z_j\;\partial _j- s. 
\]
The $R$-matrix of the spin-$s$ XXX chain acts in the tensor product space $%
h^{(s)}\otimes h^{(s)}.$ It is given at spectral parameter $\lambda$, by 
\cite{rmatrix}:
\beq
R_{12}^{(s)}(\lambda )=\frac{\Gamma(i\lambda -2s) \Gamma(i\lambda +2s+1)}
{\Gamma(i\lambda -J_{12}) \Gamma (i\lambda +J_{12}+1)}\;, 
\eq
where the Casimir operator $J_{12}$ satisfies the relation
\beq
J_{12}(J_{12}+1)=(\vec{S}_1+\vec{S}_2)^2=2
\vec{S}_1.\vec{S}_2+2s(s+1)\,.
\eq
One then has
\beq
H_{jk}=-i\frac d{d\lambda }\log R_{jk}^{(s=0)}(\lambda )_{\mid_{\lambda =0}}
\eq
as can be seen from the representation (\ref{hjk3}).
The matrix $R^{(s)}(\lambda )$ satisfies the Yang-Baxter equation.
This implies that 
\beq
H_n=\stackunder{j=1}{\stackrel{n}{\sum }}H_{j,j+1}
\label{ham}
\eq
belongs to an infinite set of conserved quantities 
\beq
\tau _k=-i\frac{d^k}{d\lambda^k}
\log R^{(s=0)}(\lambda)_{\mid_{\lambda=0}}\;,\;k=0,1,2,...
\eq
These quantities are simultaneously diagonalized by means 
of the algebraic Bethe
Ansatz. In this approach one defines the Lax operators,
\[
L_k^{(s)}(\lambda )=\lambda I_k\otimes I+i\vec{S}_k\ot 
\vec{\sigma }=\left( 
\begin{array}{ccc}
\lambda +iS_k^3 &  & iS_k^{-} \\ 
&  &  \\ 
iS_k^{+} &  & \lambda -iS_k^3
\end{array}
\right) 
\]
and the monodromy and transfer matrices:
\beq
T_a(\lambda )=L_n(\lambda )...L_1(\lambda )=\left( \begin{array}{cc}
A(\lambda ) & B(\lambda)\\
C(\lambda ) & D(\lambda )\end{array} \right)
\,,\;\Lambda (\lambda ) 
=\hbox{tr}\, T_a(\lambda )=A(\lambda )+D(\lambda )
\eq
The conserved quantities 
\[
\hat{q}_{n-k}=\frac{1}{k!}\frac{d^k}{d\lambda^k}
\Lambda (\lambda )_{\mid_{\lambda=0}}\;,\;\;k=1,...,n-2
\]
commute with the operators $\tau _k$ because of the Yang-Baxter equation.
One obtains
\[
\Lambda (\lambda )=2\lambda^n+\hat{q}_2\lambda^{n-2}+
\hat{q}_3\lambda^{n-3}+\cdots +\hat{q}_n, 
\]
where
\[
\hat{q}_k=\sum_{n\geq j_1>...>j_k\geq 1}i^k z_{j_1 j_2}
z_{j_2 j_3}...\, z_{j_k j_1}\partial_{j_1}...\,\partial_{j_k}. 
\]
In particular we have the Casimir operator of the conformal algebra:
\[
\hat{q}_2=\sum_{n\geq j>k\geq 1}z_{j k}^2\partial_j\partial_k
=-S^3S^3-\frac{1}{2}\left( S^+ S^- + S^- S^+ \right) \equiv -h(h-1)\;. 
\]

A set of simultaneous eigenvectors of the $\hat{q}_k$ is also
a set of simultaneous eigenvectors of the $\tau_k.$ Such vectors are then
given by the algebraic Bethe Ansatz approach. A subclass can be
written as:
\begin{equation}
\varphi _{n,\{\lambda_i\}}^{(s=0)}\left( \left\{ z_j\right\} ;0\right)
=z_{12}z_{23}...z_{n1}B^{(s=-1)}(\lambda_1)...B^{(s=-1)}(\lambda_p)
\frac{1}{z_1^2...z_n^2}
\label{evec1}
\end{equation}
The appearance of the $B$-operators for spin $-1$ is due to 
the existence of a trivial highest-weight vector for the spin $0$ chain. 
The spin $-1$ chain is related
by a similarity transformation to the spin $0$ chain. In particular the
eigenvalues $q_k^{(s=0)}$ and $q_k^{(s=-1)}$ are equal.
The parameters $\{\lambda_i\}$ satisfy the Bethe Ansatz equations for
spin $-1$:
\begin{equation}
\left( \frac{\lambda _k -i}{\lambda _k +i}\right) ^n=-\prod_{j=1}^p
\frac{\lambda _k-\lambda _j+i}{\lambda _k-\lambda _j-i}\;,\;\;\; k=1,....,p\,.
\label{bae}
\end{equation}
The eigenvalues are given by
\begin{equation}
\varepsilon_n=-2n-2\sum_{j=1}^p\frac{1}{\lambda_j^2+1}\;.
\label{eval1}
\end{equation}

In fact it is possible to reformulate equations (\ref{bae}), (\ref{eval1}) 
and (\ref{evec1}) as: 
\bar
\label{beq}
\Lambda^{(s=-1)}(\lambda )Q_n(\lambda)&=&(\lambda +i)^n
Q_n(\lambda +i)+(\lambda -i)^n Q_n(\lambda -i)\;,\\
\varepsilon_n &=&-2n+i\left( \frac{Q_n^{\prime}(-i)}{Q_n(-i)}-
\frac{Q_n^{\prime} (i)}{Q_n(i)}\right) , 
\label{eval2}\\
\varphi _{n,\{\lambda_j\}}^{(s=0)}\left(\{z_j\};0\right)&=&
z_{12}z_{23}...
z_{n1}(i S^-)^{h-n}Q_n(x_1)...Q_n(x_{n-1})\frac{1}{z_1^2...z_n^2}
\label{evec2}
\end{eqnarray}
Equation (\ref{beq}) is the eigenvalue version of the Baxter equation;
$\Lambda^{(s=-1)}(\lambda )$ and $Q_n(\lambda)$ stand for the operators and
their eigenvalues.
In this approach one looks for a solution $Q_n(\lambda )$ {\em analytical} 
in {\bf C}. In (\ref{evec2}), $(x_1,...,x_{n-1})$ are operators
such that
\[
B(\lambda)=iS^-(\lambda -x_1)...(\lambda -x_{n-1})\,,\;\;\; S^{-}=-
\sum_{j=1}^n \partial_j\,.
\]

For a {\em polynomial} solution $Q_n(\lambda )=\prod_{j=1}^p
( \lambda -\lambda_j)$ of the Baxter equation we
obtain eqs. (\ref{evec1}), (\ref{bae}) and (\ref{eval1}). Thus the system 
(\ref{beq}), (\ref{eval2}) and (\ref{evec2}) gives a larger class of solutions
than the class of polynomial solutions.

We conclude by rewriting the conformal invariance property using
the generators $S^{\pm},S^3$ of the whole chain, for spin zero:
\begin{eqnarray*}
\left( \stackunder{j=1}{\stackrel{n}{\sum }}\partial _j+\partial _0\right)
\chi _{n\left\{ q\right\} }=\left( \stackunder{j=1}{\stackrel{n}{\sum }}%
z_j\partial _j+z_0\partial _0+h\right) \chi _{n,\left\{ q\right\} } \\
=\left( \stackunder{j=1}{\stackrel{n}{\sum }}z_j^2\partial _j+z_0^2\partial
_0+h\right) \chi _{n,\left\{ q\right\} }=0\,,
\end{eqnarray*}
and similarly with anti-holomorphic operators.
This can be rewritten as
\begin{eqnarray*}
S^3 \chi_{n,\left\{ q\right\} }\left( \left\{ z_j,\overline{z}
_j\right\} ;0,0\right) &=&-h \,\chi_{n,\left\{ q\right\} }\left( \left\{
z_j,\overline{z}_j\right\} ;0,0\right) \\
S^+ \chi_{n,\left\{ q\right\} }\left( \left\{ z_j,\overline{z}_j\right
\} ;0,0\right) &=&0 \\
\chi_{n,\left\{ q\right\} }\left( \left\{ z_j,\overline{z}_j\right\} ;z_0,
\overline{z}_0\right) &=& \chi_{n,\left\{ q\right\} }\left( \left\{
z_{j0},\overline{z}_{j0}\right\} ;0,0\right) .
\end{eqnarray*}
Therefore $(-h)$ can be identified as the total spin of the vector 
$\chi_{n,\left\{ q\right\} }$ and one must also have $-h=-1.n-p$, and thus 
$p=h-n$. 
This implies $h=n,n+1,...$ for the polynomial solutions. There is also a $%
h\rightarrow 1-h$ symmetry in the problem which allows us to consider 
only Re$\, h\geq \frac 12$.

\section{The General Solution of the Baxter Equation}\label{diff}

Following refs. \cite{fk,k} we set 
\begin{equation}
Q_n(\lambda )=\int_C\frac{dz}{2\pi i}z^{-i\lambda -1}(z-1)^{i\lambda -1}
\tilde{Q}_n(z) 
\label{mel}
\end{equation}
where the closed path $C$ encircling the two points 0 and 1 counterclockwise
is such that the integrand is uniform. The
Baxter equation becomes an $n^{th}$ order differential equation for 
$\tilde{Q_n}(z)$:
\begin{equation}
\left[ \left( z(1-z)\frac d{dz}\right)^n+z(1-z)\sum_{k=0}^{n-2}i^{n-k}q_{n-k}
\left( z(1-z)\frac 2{dz}\right)^k\right] \tilde{Q}_n(z)=0\,.
\label{dif}
\end{equation}
This is a Fuchsian differential equation with three regular singular points, 
$z=0,1$ and $\infty$ \cite{ince}. The indicial equations for
these singular points are 
\[
s^n=0\;\; \hbox{for}\;\; z=0 \;\;\hbox{and}\;\; z=1\,, 
\]
\[
s(s-1)...(s-n+3) \left[ (s-n+2)(s-n+1)+q_2\right] 
=0\;\; \hbox{for}\;\; z=\infty . 
\]

One can then look for logarithm-free solutions in the form of an entire
series around any of the three singular points. We consider a series
expansion around the point $z=0$ of the form:
\begin{equation}
\tilde{Q}_n(z)=\sum_{k\geq 0}a_k z^k  
\label{exp}
\end{equation}
It is then easy to see that the sequence $\left\{ a_k\right\} $ will satisfy
an $n$-term linear recurrence relation of the form
\begin{equation}
\sum_{i=0}^{n-1}a_{k+i}\, p_i(k+i)=0\,,
\label{rec}
\end{equation}
where
\[
p_{n-1}(k)=k^n\,,\; p_0(k)=k(k+1)...(k+n-3)\left(
(k+n-2)(k+n-1)+q_2\right)\, , 
\]
and for all $i$, $p_i(k)$ is a polynomial of degree $n$ in $k$
with a linear dependence on the parameters $\{q_k\}$.

Expansions around $z=1$ and $z=\infty$ lead to similar
recurrence relations. An expansion on the basis of the Legendre polynomials
was considered in ref.~\cite{k,lipleg}.

Deforming the integration contour in eq.~(\ref{mel}), one can rewrite 
$Q_n(\lambda)$ as
\beq
Q_n(\lambda )=i\frac{\sinh (\pi\lambda)}\pi \int_0^1
x^{-i\lambda -1}(1-x)^{i\lambda -1}\tilde{Q}_n(x)dx
\label{mel1}
\eq
Plugging the series expansion (\ref{exp}) into eq.~(\ref{mel1}) one obtains:
\begin{equation}
Q_n(\lambda )=\sum_{k=1}^\infty \frac{a_k}{(k-1)!}\prod_{l=1}^{k-1}(l -
i\lambda)
\label{sol}
\end{equation}
Thus a polynomial solution of degree $p$ in $z$ of the differential
equation (\ref{dif}) yields a polynomial solution, of degree $p-1$ in
$\lambda$, to the Baxter equation. It is possible to verify directly that,
at least for the first few values of $n$, the expression (\ref{sol}) provides a
solution of the Baxter equation if the recurrence (\ref{rec}) is satisfied.
To do 
so one just expands the polynomials $\lambda^j$ and $(\lambda \pm i)^n$ 
in a way
which is compatible with an 
expansion over the basis provided by 
\[
P_0(\lambda) =1\,,\;\; P_k(\lambda ) =\prod_{l=1}^{k-1}(l -i\lambda )\;\,,\;\;
k=1,2,...
\]

We note here that for any periodic function $f$ of $\lambda$
with period $i$, one can define a new solution of the Baxter equation by
multiplying any solution with $f(\lambda )$.

We now briefly address the issue of the convergence of a series of the
type (\ref{sol}). For $n=2$ the solution recessive at 0 of the differential
equation (\ref{dif}) is the hypergeometric function
\[
\tilde{Q}_2(z)=\, _2F_1(h,1-h;1;z) 
\]
for $h\notin$ {\bf Z}. For $k$ large, $a_k\sim\frac{1}{k}+{\cal O}
\left( \frac 1{k^2}\right) $ and the series (\ref{sol}) 
converges absolutely and
uniformly for Im$\,\lambda <0$, where $Q_2(\lambda )$ is therefore
analytical. The Baxter equation allows to extend analytically $Q_2(\lambda)$
to {\bf C}$\backslash i${\bf N}.

There is generically an infinite number of poles at the points
$0,i,2i,...$ By calculating 
\[
Q_2(\lambda )(2\lambda^2+q_2) -Q_2(\lambda -i)(\lambda -i)^2\,,
\]
and its derivative at $\lambda =0$, with the help of the recurrence
relation, we find that 
the pole at 0 is simple with residue $\frac{-i}{(h-1)!(-h)!}\,.$ The Baxter
equation implies then 
the existence of other simple poles at $i,2i,...$

We also found numerically that $Q_2(\lambda )$ increases exponentially as 
$\lambda \rightarrow -i\infty$ for $ h=1/2$ and $h=5/2$;
we believe that this is the case for generic value of $h$.
If one assumes $Q_2(\lambda )\sim \lambda^\alpha$ as $\lambda$ tends to 
{\em real} infinity then the Baxter equation implies that 
$\alpha =h-2$.\footnote{
Note however that an analytical function in {\bf C} 
cannot have a branch-type singularity
of the form $\lambda^{h-n} (h\notin$ {\bf Z}) as $\lambda$ tends to
complex infinity.}
Numerical trials confirm this assumption.

The other solution of the differential equation, $_2F_1(h,1-h;1;1-z)$,
gives 
the series $Q_2(-\lambda )$ which is therefore also a solution of the Baxter
equation with the same value of $q_2$. One way to construct a solution of
the Baxter equation analytical in {\bf C} is to consider
\[
\sinh(2\pi \lambda )\left( c_1 Q_2(\lambda )+c_2 Q_2(-\lambda )\right) 
\]
for any values of $c_1$ and $c_2$. The behaviour of this analytical 
solution at real infinity is not
likely to be $\lambda^{h-2}$ however. In this respect 
we seem to disagree with the results of \cite{fk,k}.

We now consider $n=3$, the first ``non-trivial'' case. The recurrence 
(\ref{rec}) becomes:
\bar
a_{k+2}(k+2)^3 - a_{k+1}\left[ i q_3+(k+1)\left( q_2+(2 k+3)(k+2)\right)
\right] \label{rec3}\\
+a_k k\left( (k+1)(k+2)+q_2\right) =0\;\;
\hbox{for}\;\; k=-1,0,1,...,\;\;\hbox{with}\;\; a_{-1}\equiv 0\,.\nonumber
\ear
It does not seem possible to find an explicit solution to this
recurrence.
However, because the indicial equations at $z_0=0$ and $z_0=1$ are 
$s^3=0$, the general solution of the
differential equation around these points is of the type $f_1(z)+f_2(z)
\log (z-z_0)+f_3(z)\log^2(z-z_0)$ where
$f_i(z)$ are regular functions at $z=z_0$. The only possible
singularities of the solutions are the three singular points. Thus
$z=1$ is the only singularity on the circle of radius one of the 
solution (\ref{exp}). The 
logarithmic behaviour of the solution around one 
implies then, using the Darboux theorem, the following behaviour of 
$a_k$ as $k$ tends to infinity:
\beq
a_k=\left( \frac{\alpha _1}k+\frac{\alpha _2}{k^2}+\cdots\right)\log k +\frac{
\beta _1}k+\frac{\beta _2}{k^2}+\cdots
\label{asym}
\eq
From the recurrence equation, it is possible to find relations between the 
coefficients  appearing in (\ref{asym}).
Numerical trials
confirm this logarithmic behaviour for $ka_k$ as $k$ becomes large.
Such a behaviour allowed us to show that for Im$\,\lambda < 0$
the series (\ref{sol}) converges uniformly. One then extends analytically this
solution to almost all the complex plane through the Baxter equation. Again,
proceeding similarly to the $n=2$ calculation, the points $i${\bf N} turn
out to be poles. They are of order two.
For instance we find for $\lambda$ near 0
\[
Q_3(\lambda)\sim -\frac{\alpha_1}{\lambda^2}\;.
\]
For the solution $Q_{3\left\{
q_2,q_3,\right\} }(\lambda )$ just defined one then considers the function
$Q_{3\left\{ q_2,-q_3\right\} }(-\lambda ).$ This is also a solution of the
Baxter equation with $(q_2,q_3)$ as parameters and
\beq
\sinh^2(2\pi\lambda)\left( c_1 Q_3\{q_2,q_3\}(\lambda )+
c_2 Q_3\{q_2,-q_3\} (-\lambda)\right) 
\eq
is also a solution, which is analytical in {\bf C}.
The behaviour at infinity is not likely to be simple.

For any given value of $n$, one can in principle repeat the same
analysis and expect to find poles of order $n-1$ at $i${\bf N} for the
analytically extended series solution.The approach of $\cite{vincent}$
to constructing solutions of the Baxter equation of the Toda chain
with a given behaviour at infinity, can also be tried. We are
investigating this possibility.

\section{Polynomial solutions of the Baxter Equation}\label{poly}

Polynomial solutions to the Baxter equation provide a subclass of
eigenvectors of the operators $\hat{q}_2,\hat{q}_3$ and $H_n$.

It is possible to obtain polynomial solutions from the series (\ref{sol}) by
requiring the sequence $\{a_k\}$ to truncate. For a polynomial of
degree $p$ it is necessary to have $a_{p+i}=0$ for all $i\geq 2$.
However, for a given $n$, the sequence satisfies the $n$-term recurrence
relation (\ref{rec}) and it is enough to require
\[
a_{p+1}\neq 0\;\;\hbox{and}\;\;a_{p+2}=a_{p+3}=\cdots =a_{p+n}=0\,.
\]
These relations are enough to quantize the eigenvalues $q_2,...,q_n$. For
instance setting $k=p+1$ in (\ref{rec}) one obtains
\begin{equation}
q_2=-(p+n)(p+n-1)=-2\left(\begin{array}{c}
p+n \\ 
2
\end{array}
\right)  
\label{q2}
\end{equation}
Writing $q_2=-h(h-1)$ gives $h=p+n$ as was found in \cite{fk,k}.
The conditions on $q_3,...,q_n$ are given by the roots of a coupled system
of polynomial equations in the variables $q_i$'s.
For $n=3$ the discretized values of $q_3$ are given by the roots of a degree 
$p+1$ polynomial.\footnote{{\it Mathematica }and {\it Maple} give exact roots
up to $h=15$, and fail otherwise.} It turns out that all the roots seem to
be real, in accordance with the results of \cite{fk,k}.

We now derive equations for the invariants of polynomial solutions to the
Baxter equation. Let
\begin{eqnarray}
\sigma_q(X_i) &\equiv &\sigma_q (X_1,...,X_p)\equiv \sum_{1\leq
r_1<r_2<\cdots <r_q\leq p}\!\!\!\!\!\!\!\!\! X_{r_1}...\, X_{r_q}\,,\;\;
q=1,...,p, \\
\sigma_0 (X_i) &\equiv &1
\end{eqnarray}
be the elementary symmetric polynomials.
The polynomials $\sigma_q (X_i)$ satisfy the relations
\begin{equation}
\sigma_q(X_i+a)=\sum_{r=0}^q\left( 
\begin{array}{c}
p-q+r \\ 
p-q
\end{array}
\right) a^r\sigma_{q-r}(X_i)  
\label{trans}
\end{equation}

A polynomial solution $Q_n(\lambda)$ of the Baxter equation, whose roots
therefore satisfy the Bethe Ansatz equations, can be written:
\[
Q_n(\lambda )=\prod_{j=1}^p (\lambda -\lambda_j)=\sum_{j=0}^p (-1)^{p-j}
\sigma_{p-j}\lambda^j
\]
where
\[
\sigma _j\equiv \sigma _j(\lambda _1,...,\lambda _p)\,.
\]
We then rewrite the Baxter equation as a set of equations:
\begin{equation}
\sum_{k=0}^{n-2}(-1)^{n-1+k}q_{n-k}\sigma _{p-m+k}=2\!\!\!\!\sum_{k=0}^{\left[ 
\frac{p-m+n-2}{2}\right] }(-1)^k\left( 
\begin{array}{c}
m+2k+2 \\ 
m
\end{array}
\right) \sigma_{p-m+n-2-2k}\,,
\label{lin}
\end{equation}
\begin{eqnarray*}
0 &\leq &m\leq p+n-2\,,\qquad \sigma_{-n+2}=\cdots =\sigma_{-1}=0\,, \\
\sigma_0 &=&1\,,\;\; \sigma_{p+1}=\cdots =\sigma_{p+n-2}=0.
\end{eqnarray*}
where $[x]$ is the integer part of $x$. The $p$ equations,
$m=1,...,p$, give a triangular linear system, with parameters $(q_3,...,q_n)$,
for the unknowns $(\sigma_1,...,\sigma_p)$. The equation $m=p+n-2$ gives,
as expected, equation (\ref{q2}). The $n-2$ remaining equations, $m=0$ and
$m=p+1,...,p+n-3$, give the quantization conditions, once the $\sigma
_k$'s are solved for as polynomials in the parameters 
$(q_3,...,q_n)$.

For $n=2$, we get that the polynomial $Q_2(\lambda )$ is even (odd) for $p$
even (odd). For $n=3$, the quantization condition for $q_3$  can be written as
a vanishing determinant of a matrix with a lower triangular part and one
non-vanishing line above the diagonal, once $\sigma_0$ is introduced in
the set of unknowns.\footnote{We checked using {\it Mathematica} that,
for $p$ up to 15, the values of $q_3$ coincide with those obtained from the
recurrence relation for $n=3$.}

The energy $\varepsilon_n$ of the Hamiltonian (\ref{ham}) can be rewritten as
follows:
\bar
\varepsilon_n &=&-2n+i\left( \frac{Q_n^{\prime}(-i)}{Q_n(-i)}-\frac{
Q_n^{\prime }(i)}{Q_n(i)}\right)\label{e1}\\
&=&-2n-2\sum_{k=1}^p\frac 1{\lambda_k^2+1} \label{e2}\\
&=&-2n+i\left( \frac{\sigma_{p-1}(\lambda_k -i)}{\sigma_p(\lambda_k -i)}
-\frac{\sigma_{p-1}(\lambda_k +i)}{\sigma_p(\lambda_k+i)}\right) . 
\label{e3}
\ear
One can then use relation (\ref{trans}) to rewrite $\varepsilon_n$ in terms of
the $\sigma_k$'s.
Equations (\ref{lin}) provide a {\em new } way for looking at the Bethe Ansatz
equations. It is also possible to obtain similar equations for any spin $s$,
not just the case $s=-1$ under consideration. 

\section{A new expression for the eigenvectors}\label{vectors}

In the framework of the algebraic Bethe Ansatz, the eigenvectors are
obtained as the repeated action of a ``lowering'' operator $B(\lambda)$ on
a highest weight state, or pseudo-vaccum state, as in equations
(\ref{evec1}) and (\ref{evec2}). However both expressions are unwieldy 
as the number of Bethe Ansatz roots increases, 
or for non-integer values of $h$. We now
develop compact expressions for the eigenvectors as linear combinations
of eigenvectors of the operator $\hat{q}_2$.

Define the vectors
\begin{equation}
\varphi_{\alpha_1,\alpha_2,\alpha_3}(z_1,z_2,z_3)\equiv z_1^{\alpha_1-h}
z_2^{\alpha_2-h}z_3^{\alpha_3-h}z_{12}^{\alpha_3}z_{23}^{\alpha_1}
z_{31}^{\alpha_2}\,,
\label{fia}
\end{equation}
where
\beq
\af_1\,,\; \af_2\,,\; \af_3\; \in \hbox{\bf C}
\,,\;\; \af_1 +\af_2 +\af_3 =h \,.
\label{haf}
\eq
In what follows we always assume (\ref{haf}) to hold. 
The form of the vector $\varphi_{\alpha_1,\alpha_2,\alpha_3}(z_1,z_2,z_3)$ 
is suggestive of a
generalization of the pomeron $(n=2)$ eigenvectors:
\beq
\varphi_h(z_1,z_2;z_0)=\left( \frac{z_{12}}{z_{10}z_{20}}\right)^h\,.
\label{pome}
\eq
One can directly verify that the functions (\ref{fia})
satisfy the conformal invariance
property (\ref{confo}), or by viewing it as a four-point function of a
conformally invariant theory \cite{polya}. Such four-point
functions were considered by Lipatov in a different approach 
\cite{liprev}. These functions  are eigenvectors of
\[
\hat{q}_2 = \sum_{j=1}^3 z_{ii+1}^2 \partial_i\partial_{i+1}\,,
\]
with eigenvalues $-h(h-1)$.
One also has the linear dependence relations
\begin{equation}
\varphi_{\alpha_1-1,\alpha_2+1,\alpha_3}(z_i)+\varphi_{\alpha_1-1,\alpha_2,
\alpha_3+1}(z_i)
=-\varphi_{\alpha_1,\alpha_2,\alpha_3}(z_i)
\label{lc}
\end{equation}
with two other equivalent relations where $\af_1-1\rightarrow \alpha_2-1$ 
or $\alpha_1-1\rightarrow \alpha_3-1$.
Other more complicated relations exist. Note that (\ref{lc}) 
is satisfied without the constraint (\ref{haf}).

The action of $\hat{q}_3$ on $\varphi_{\alpha_1,\alpha_2,\alpha_3}(z_i)$  is
a tedious but straightforward calculation. We obtain:
\bar
i\hat{q}_3 \varphi_{\alpha_1,\alpha_2,\alpha_3}(z_i) &=&
\left( \af_3\af_2(\af_2-1)+\af_2\af_1(\af_1-1)\right.\nonumber\\
&+&\left.\af_1\af_3(\af_3-1)\right)
\varphi_{\alpha_1,\alpha_2,\alpha_3}(z_i) \nonumber\\
&-&\af_1(\af_1-1)(\af_1-h)\varphi_{\alpha_1-1,\alpha_2+1,\alpha_3}(z_i)
\nonumber\\
&-&\af_2(\af_2-1)(\af_2-h)\varphi_{\alpha_1,\alpha_2-1,\alpha_3+1}(z_i)
\nonumber\\ 
&-&\af_3(\af_3-1)(\af_3-h)\varphi_{\alpha_1+1,\alpha_2,\alpha_3-1}(z_i)
\label{q3}
\ear
The action of $\hat{q}_3$ is that of a step operator. It is
therefore natural to look for simultaneous eigenvectors of $\hat{q}_2$ 
and $\hat{q}_3$ (and therefore of $H_3$)
as linear combinations of the eigenvectors of $\hat{q}_2$,
that is, as a kind of coherent states.

The eigenvalues of $\hat{q}_2$ are highly degenerate:
for a fixed value of $h$, two complex parameters label this degeneracy. In
order to control such degeneracy, we first consider
the plane determined by 
(\ref{haf}) in {\bf C}$^3$, and fix 
a point $(\alpha_1,\alpha_2,\alpha _3)$ in it.
The set 
\[
\varphi_{\alpha_1,\alpha_2+m,\alpha_3-m}\,,\;\;\varphi_{\alpha_1-n,\alpha_2,
\alpha_3+n}\,,\;\;\varphi_{\alpha_1+p,\alpha_2-p,\alpha_3}\,,\;\; m,n,p\in 
\hbox{\bf N}\,,
\]
is a basis for the space
\[
\varphi_{\alpha_1+m,\alpha_2+n,\alpha_3+p}\,,\;\; m,n,p\in 
\hbox{\bf Z}\,,\;\; m+n+p=0\,.
\]
This space can be represented as shown in figure 1, where the vertices
correspond to the foregoing vectors. The vertices on the three solid lines
are the basis vectors. The triangle in bold lines represents a three-term
linear dependence relation of the kind (\ref{lc}).
This diagram is reminiscent of the $sl(3)$ weight lattice.
\begin{figure}[htbp]
\centering
\epsfysize=6.0cm{\centerline{\epsfbox{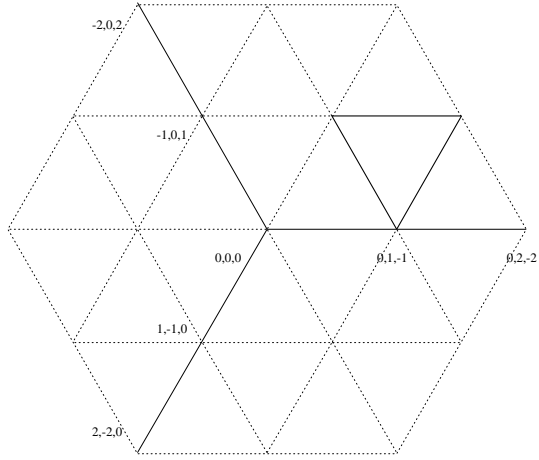}}}
\caption[Visualization of the space with origin $(\alpha_1,\alpha_2,
\alpha_3)$, represented as $(0,0,0)$.]
{\label{ba} Visualization of the space with origin $(\alpha_1,\alpha_2,
\alpha_3)$, represented as $(0,0,0)$. }
\end{figure}

We can then look for eigenvectors of $\hat{q}_3$ as a
linear combination of basis vectors,
\[
\phi_{q_3}(z_i)=\sum_{m\geq 0}a_m\varphi_{\alpha_1,\alpha_2+m,\alpha_3-m}
+\sum_{n\geq 0}b_n\varphi_{\alpha_1-n,\alpha_2,
\alpha_3+n}+\sum_{p\geq 0}c_p\varphi_{\alpha_1+p,\alpha_2-p,\alpha_3}\,.
\]
Requiring $\phi_{q_3}(z_i)$ to be an eigenvector of $\hat{q}_3$
gives a complicated set of coupled recurrence relations with an infinite
number of terms. To obtain a more manageable set of relations we restricted
the point $(\alpha _1,\alpha _2,\alpha _3)$ to one of the three points 
$(h,0,0)$, $(0,h,0)$ and $(0,0,h)$. The recurrence relations then decouple
and become  three-term recurrence relations of the type already  encountered
in section \ref{diff}. We now give these relations for the point $(h,0,0)$.
We note that for integer $h$ it is enough to consider this point.
Consider the vector
\beq
\phi_{h,q_3}(z_i)= \sum_{m\geq 1}a_m\varphi_{h,m,-m}(z_i)+
\sum_{m\geq 1}b_m\varphi_{h-m,0,m}(z_i)+
\sum_{m\geq 1}c_m\varphi_{h+m,-m,0}(z_i) \,.
\label{anz}
\eq
We have not considered the three $m=0$ terms in this sum because such terms
are in the kernel of $\hat{q}_3$ and we are interested in
non-vanishing eigenvalues. The action of $\hat{q}_3$ on 
(\ref{anz}) is easy to obtain from (\ref{q3}). We get three uncoupled
recurrence relations for the coefficients in the sum which means that one
can consider each sum by itself as an eigenvector. The recurrence equations
are:
\bar
(m+1)m(m+1-h)a_{m+1}&+&\left( iq_3+m(2m^2-h(h-1))\right)a_m\nonumber\\
&+&m(m-1)(m+h-1)a_{m-1}=0\;,\label{recs1}\\
(m+1)m(m+1-h)b_{m+1}&+&(i q_3 +m(m-h)(2m-h))b_m\nonumber\\
&+&(m-1)(m-h)(m-1-h)b_{m-1}=0\,,\label{recs2}\\
(m+1)(m+1+h)(m+h)c_{m+1}&+&(i q_3+m(m+h)(2m+h))c_m\nonumber\\
&+&m(m-1)(m-1+h)c_{m-1}=0\;. \label{recs3}
\ear
These recurrence relations are different from eq.~(\ref{rec3}) but 
of the same type. Polynomial solutions of the Baxter equation 
correspond to finite sums in (\ref{anz}). For instance, the solutions
\[
Q_3(\lambda)=\lambda\pm\frac{1}{\sqrt{3}}
\]
with $q_3=\pm 2\sqrt{3}$ and $q_2=-12$ ($h=4$), yield the two eigenvectors:
\bar
\label{ex1}
\phi_{4,\pm 2\sqrt{3}}(z_i)&=&\mp\sqrt{3}\varphi_{121}+i(\varphi_{112}-
\varphi_{211})\\
&=& \mp\sqrt{3}(\varphi_{103}+\varphi_{301}+2\varphi_{202})+i(\varphi_{301}
-\varphi_{103})\;.
\label{ex2}
\ear
The first expression is calculated directly from (\ref{evec1}), where
\begin{eqnarray*}
B^{(-1)}(\lambda)&=&i\lambda^2(S_1^-+S_2^-+S_3^-)+\lambda(S_2^-S_1^3-S_2^3S_1^-
+S_3^-S_1^3-S_3^3S_1^-\\
\hspace{-1cm}&+&S_3^-S_2^3-S_3^3S_2^-)
-i(S_3^-S_2^+S_1^-+S_3^-S_2^3S_1^3 +S_3^3S_2^3S_1^--S_3^3S_2^-S_1^3)
\label{bl}
\end{eqnarray*}
with spin $-1$ generators.

Equation (\ref{ex2}) is obtained from equation (\ref{ex1}) by 
using the relations
(\ref{lc}). It is also obtained from (\ref{anz}) and the recurrence relation
(\ref{recs2}).
More generally, the relations (\ref{recs1}--\ref{recs3}) 
can be truncated
to a finite number of terms by
requiring $h=-n_0$ for (\ref{recs1}) and (\ref{recs3}), 
$h=n_0+1$ for (\ref{recs2}) ($q_2=-h(h-1)=-n_0(n_0+1)$ for the three cases), 
and  the relations for $n=n_0$ to hold. 
We verified that the discretized values of $q_2$ and $q_3$
we obtain are the same as the values we found in sections 
3 and 4, for many values of $h$. 
We also found an additional $q_3=0$
solution for the recurrences (\ref{recs1}) and (\ref{recs2}).

Eigenvectors common to $\hat{q}_2$ and $\hat{q}_3$ 
and with vanishing eigenvalue $q_3$ are just linear combinations of
pomeron eigenvectors. This becomes clear if one solves the partial
differential equation
\[
\hat{q}_3 \varphi =i z_{12}z_{2}z_{31}\partial_1\partial_2\partial_3
\varphi=0\,;
\]
it implies that $\varphi (z_i)$ is a sum of three functions which depends on
only two of the three variables $(z_1,z_2,z_3)$. For instance one has :
\[
\varphi_{h,0,0}(z_{10},z_{20},z_{30})=\varphi_h(z_2,z_3;z_0)\,,\;
\varphi_{111}(z_{i0})=\frac{1}{3}(\varphi_{003}+\varphi_{030}+\varphi_{300})
(z_{i0})\,.
\]
The action of $H_3$ on $\varphi_{q_3=0}$ is straightforward because of
the foregoing remarks:
\[
H_3\varphi_{q_3=0}=H_2(q_2)\varphi_{q_3=0}\,.
\]
However the action of $H_3$ on $\varphi_{q_3\neq 0}$ is not simple.
For polynomial solutions, corresponding to truncated sums in (\ref{anz}), we
obtain the eigenvalue of $H_3$ from (\ref{e1}), (\ref{e2}) or (\ref{e3}).

The Ansatz (\ref{anz}) covers all the polynomial solution class of the Baxter
equation. We also believe it provides a space rich enough to cover most of
the non-polynomial solutions we are interested in.

An extension of the foregoing approach to a chain with four sites and larger
sizes can be considered. For four sites, the ``elementary'' functions, which
are eigenvectors of $\hat{q}_2$ with eigenvalue $q_2 =-h(h-1)$, are:
\begin{equation}
\varphi_{q_2,\{\alpha\}}(z_i)=z_1^{\alpha_1-h}z_2^{\alpha_2-h}
z_3^{\alpha_3-h}z_4^{\alpha_4-h}z_{12}^{\alpha_{34}-h}
z_{23}^{\alpha_{14}-h}z_{34}^{\alpha_{12}-h}z_{41}^{\alpha_{23}-h}
z_{13}^{\alpha_{24}-h}z_{24}^{\alpha_{13}-h}\,,
\label{elem4}
\end{equation}
with 
\begin{eqnarray*}
h=\af_{12}+\af_{13}+\af_{14}+\af_{23}+\af_{24}+\af_{34}\,,\\
\af_1=h-\af_{23}-\af_{24}-\af_{34}\,,\;\;
\af_2=h-\af_{13}-\af_{14}-\af_{34}\,,\\
\af_3=h-\af_{12}-\af_{14}-\af_{24}\,,\;\;\af_4=h-\af_{12}-\af_{13}-\af_{23}\,.
\end{eqnarray*}
The actions of $\hat{q}_3$ and  $\hat{q}_4$ can be calculated, 
and these operators are then seen to act as step
operators. One can then take an Ansatz for the eigenvectors and find the
recurrence relations for the coefficients. The generalization of the
elementary functions (\ref{elem4}) to larger chains is straightforward.

\section{Conclusion} 
The determination of the ground state of the Hamiltonian $H_n$ for $n\geq 3$
is still an open problem.  In this paper we analyzed the Baxter 
equation connected with the
diagonalization. Common features emerged such as the series solutions and
the $n$-term recurrence associated with them. It seems impossible to reduce
such recurrences to simpler ones. Such recurrence relations emerged again
when we looked for a simple form for the eigenvectors. We also recast the
Baxter equation as a linear system suitable for the search for
polynomial solutions.

\bigskip\ {\bf Acknowledgements:} It is a pleasure to thank V.\
Pasquier, M.\ Bauer, R.\ Peschanski and S.~Nonnenmacher
for many useful discussions.
We would like to thank G.\ Korchemsky for fruitful 
discussions. S.\ W. is grateful
to L.\ N.\ Lipatov and J.\ Bartels for discussions.
Special thanks to J.\ B.\ Zuber for bringing to our attention the deep $
\frac{SW}{ZM}$ mirror symmetry present in this problem, and for a careful
reading of the manuscript.

\end{document}